\documentclass[%
reprint,
amsmath,amssymb,
aps,
]{revtex4-1}

\usepackage{graphicx}
\usepackage{dcolumn}
\usepackage{bm}

\begin{document}

\preprint{APS/123-QED}

\title{Highly Non-Degenerate Two-Photon Absorption in Silicon Wire Waveguides}

\author{Nicolas Poulvellarie}
\affiliation{%
 OPERA-Photonique, Universit\'e libre de Bruxelles (ULB), 50 Av. F. D. Roosevelt, CP 194/5, B-1050 Bruxelles, Belgium\\
}%
\author{Charles Ciret}
\affiliation{%
Laboratoire de Photonique d'Angers EA 4464, Universit\'e d'Angers, 2 Bd Lavoisier, 49000 Angers, France
 }%
\author{Bart Kuyken}
\affiliation{
Photonics Research Group, Department of Information Technology, Ghent University-IMEC, Ghent B-9000, Belgium\\
}%
\author{Fran\c cois Leo}%
\author{Simon-Pierre Gorza}%
\email{sgorza@ulb.ac.be}
\affiliation{%
 OPERA-Photonique, Universit\'e libre de Bruxelles (ULB), 50 Av. F. D. Roosevelt, CP 194/5, B-1050 Bruxelles, Belgium\\
}%

\date{\today}

\begin{abstract}
Non-degenerate two-photon absorption (TPA) is investigated in a nanophotonic silicon waveguide in a configuration such that the dispersion of the nonlinear absorption and refraction cannot be neglected. It is shown that a signal wave can strongly be absorbed by cross-TPA by interaction with a low energy pump pulse (1.2\,pJ), close to the half-bandgap and experiencing low nonlinear absorption. The experiments are very well reproduced by numerical simulations of two-coupled generalized nonlinear Schr\"{o}dinger equations (GNLSE), validating the usual approximation made to compute the cross nonlinear coefficients in indirect-gap semiconductors. We show that the nonlinear dynamics can be well described by a single GNLSE despite the wavelength separation between the pump and the signal waves. We also demonstrate that in silicon wire waveguides and contrary to optical fibers, the dispersion of the nonlinear absorption is much larger than the dispersion of the Kerr effect. This could have an impact in the design of all-optical functions based on cross-TPA, as well as on the study of supercontinuum and frequency comb generation in integrated semiconductor-on-insulator platforms.

\end{abstract}

\pacs{Valid PACS appear here}
\maketitle


\section{\label{sec:level1}Introduction}

In the past years, silicon nanophotonic waveguides attracted a significant interest owing to its large nonlinear optical response. Nonlinear optical functionalities have been demonstrated such as all-optical signal processing\,\cite{li_error-free_2010}, wavelength conversion\,\cite{dekker_ultrafast_2006, foster_broad-band_2007} or supercontinuum generation\,\cite{hsieh_supercontinuum_2007, leo_dispersive_2014}. At telecom wavelengths, crystalline silicon however suffers from nonlinear absorption, which limits its performances. Nevertheless, the two-photon absorption (TPA) effect in silicon can be exploited for ultrashort pulse characterization\,\cite{liang_silicon_2002}, ultrafast all-optical switching\,\cite{liang_ultrafast_2005} and modulation\,\cite{moss_ultrafast_2005, mehta_all-optical_2011}, modelocking\,\cite{tien_pulse_2007} or single cycle pulse generation\,\cite{yue_uwb_2012}. Moreover, it has recently been pointed out that the nonlinear absorption plays a central role in the soliton fission mechanism at the heart of supercontinuum generation (SCG) from sub-picosecond pulses\,\cite{Ciret-18}.  

Non-degenerate two-photon absorption (ND-TPA) is the nonlinear mechanism by which two photons of different waves are simultaneously absorbed in a material. The dispersion properties of degenerate TPA in crystalline silicon are well-known\,\cite{bristow_two-photon_2007, cheng_full_2011} and have been confirmed by many experiments. However, despite the importance of non-degenerate nonlinear absorption in broadband applications, little is known about ND-TPA when the interacting wavelengths are vastly different. Recent works on ND-TPA in silicon have been reported in \,\cite{ suda_femtosecond_2009, zhang_non-degenerate_2015, sarkissian_cross_2015}. However in these studies the two interacting waves are only separated by 75\,MHz, 5\,nm and 30\,nm, respectively, and the approximation of constant TPA coefficients is relevant. In direct-gap semiconductors, contrary to indirect-gap semiconductors, ND-TPA has been extensively studied both theoretically and experimentally \,\cite{cirloganu_extremely_2011}. It is now well known that owing to the sharp absorption band-edge of direct-gap semiconductor, ND-TPA in these materials could be very large compared to its degenerate counterpart for extreme non-degenerate cases involving spectrally distant waves. Enhancement by a factor up to 1,000 has been experimentally demonstrated with applications to the detection of mid-infrared radiation mediated by a gating pulse\,\cite{cirloganu_extremely_2011,fishman_sensitive_2011}. Such an enhancement is not expected in indirect-gap semiconductors, because of the slow dependence of the linear absorption coefficient with the frequency close to the indirect bandgap energy\citep{cirloganu_extremely_2011}. The commonly adopted approximation for computing the ND-TPA or the cross phase modulation coefficients in indirect-gap semiconductors thus simply considers the total energy of the two involved photons. This approximation consists in considering the degenerate $\chi^{(3)}$ at the average frequency of the two waves, i.e. at $(\omega_p+\omega_s)/2$, instead of $\chi^{(3)}(-\omega_s;\omega_p,-\omega_p, \omega_s)$\,\cite{lin_nonlinear_2007}. As a result, photons of energy close or below the half-bandgap could experience a large cross-absorption with photons at telecom wavelengths or below. On the other hand, since the nonlinear absorption coefficient varies with the frequency, this dispersion could have a non-negligible impact on the nonlinear wave propagation, potentially larger than the dispersion of the Kerr effect.    

In this work, our aim is to experimentally investigate the ND-TPA between short pulses in indirect-gap semiconductor waveguides, sufficiently spectrally distant for the dispersion of the nonlinear absorption and refraction coefficients to have a significant impact on the propagation dynamics. Our experiment will thus serve as a test-bed for assessing the validity of the simplifying assumption which infers the cross nonlinear coefficients from their degenerate values at the average frequency. From these results we will finally discuss the ability of the single generalized nonlinear Schr\"{o}dinger equation to describe the nonlinear cross interaction between two spectrally distant waves in integrated semiconductor waveguides. Our motivation stems from the fact that this equation is commonly encountered for modelling nonlinear dynamics of broadband waves such as the generation of supercontinuum or frequency comb in these platforms.           

\section{\label{sec:level2}Experiment }

\begin{figure}[tbp]
\centering
\includegraphics[width=\linewidth]{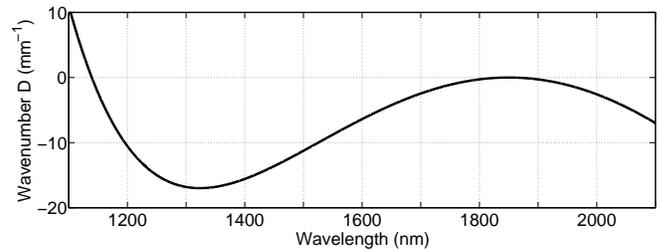}
\caption{Wavenumber $D$ of the quasi-TE fundamental mode as a function of the wavelength for a 220\,nm\,$\times$\,789\,nm silicon waveguide. The group velocity, in the reference frame of the pump at 1850\,nm, is proportional to the slope of the curve.}
\label{fig:Dispersion_curve}
\end{figure}

Integrated semiconductor wire waveguides are ideal for low power nonlinear optical functions owing to the high confinement of the light and the relatively long interaction distance with small footprint. Moreover, the geometry of the waveguides allows for the engineering of the dispersion properties which play a central role for efficient nonlinear interactions through phase matched processes. The same dispersion properties however imply that spectrally distant pulses have usually different group velocities. The resulting temporal walk-off between the pulses reduces their effective nonlinear interaction length. However, thanks to the third and/or fourth-order dispersions, two waves at very different wavelengths could have the same group velocity, allowing for longer interaction length. The dispersion of the engineered waveguide used in the experiment is shown in Fig.~\ref{fig:Dispersion_curve} for a quasi-TE mode. The dispersion parameter $D$ is defined by Eq.~(\ref{Eq_D}) and is the mode wavenumber expressed in the co-moving frame at the group velocity of a pulse at 1848\,nm. This curve shows that waves around 1850\,nm and 1300\,nm propagate with a close group velocity but in the anomalous and normal dispersion, respectively. An intense soliton pulse can thus be excited at a wavelength around 1850\,nm, while a small signal pulse around 1300\,nm will strongly interact with the soliton pulse. This is the same configuration as the one used for observing optical event horizons\,\cite{ciret_observation_2016, philbin_fiber-optical_2008}. 

\begin{figure}[b]
\centering
\includegraphics[width=7cm]{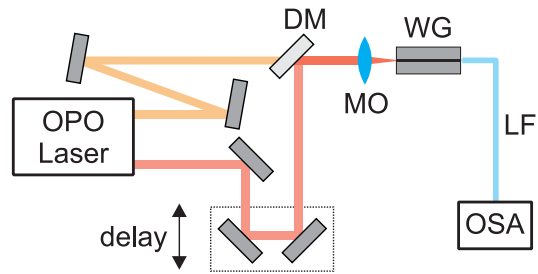}
\caption{Experimental setup. OPO: optical parametric oscillator, DM: dichroic mirror, MO: microscope objective, WG: silicon-on-insulator nanophotonic waveguide, LF: lens fiber, OSA: optical spectrum analyzer.}
\label{fig:setup}
\end{figure}

The experimental setup is shown in Fig.~\ref{fig:setup}. The wavelength tunable pump and signal pulses are generated by an optical parametric oscillator (OPO), synchronously pumped by a mode-locked Ti:sapphire laser at 82 MHz. The pump at 1848\,nm is used to excite a fundamental soliton. At this wavelength, the two-photon absorption is weaker than at telecom wavelengths. The wave around 1300\,nm acts as a signal pulse. Its power is lower than the pump power and the dominant contribution to the signal nonlinear absorption is its cross-interaction with the pump. Prior to the coupling into the silicon wire waveguide, these two beams are recombined on a dichroic mirror with a controllable delay. We consider a 2\,cm-long, 220\,nm-high, 780\,nm-wide silicon-on-insulator wire waveguide. At the waveguide output, the light is collected by means of a lens fiber and sent to an optical spectrum analyzer (either a Yokogawa AQ6370 for measuring the signal spectrum or a Fourier Transform OSA, OSA203B from Thorlabs Inc. for the pump). The lens fiber position was set to optimize the amplitude of the collected signal wave. The input pulse profiles have been characterized by an autocorrelator and an optical spectrum analyzer. We found that the pump pulses have a chirped Gaussian profile with an unchirped pulse duration of 200\,fs (FWHM) and a quadratic spectral phase profile $b*\omega^2$, with $b = 4.3\times10^{-27}$\,s$^2$ where $\omega$ is the angular frequency difference from the carrier pump frequency. The signal pulses have an hyperbolic secant profile of 152\,fs (FWHM) unchirped pulse duration and a chirp characterized by $b = 2.3\times10^{-27}$\,s$^2$.

From the waveguide dispersion curve, we can see that the wavelength at 1322\,nm is group velocity matched with the pump at 1848\,nm. For slightly detuned signals and proper initial delays, the cross phase modulation between the pump and the signal waves leads to a partial reflexion of the signal pulse in the time reference frame of the pump. Simultaneously the reflected pulse experiences a frequency shift across the group velocity match frequency, such that it preserves the wavenumber $D$~\cite{philbin_fiber-optical_2008}. The signal pulse could in principle be totally reflected by the pump pulse but we expect the ND-TPA to reduce its efficiency and affect the output spectra.

\begin{figure}[tbp]
\centering
\includegraphics[width=\linewidth]{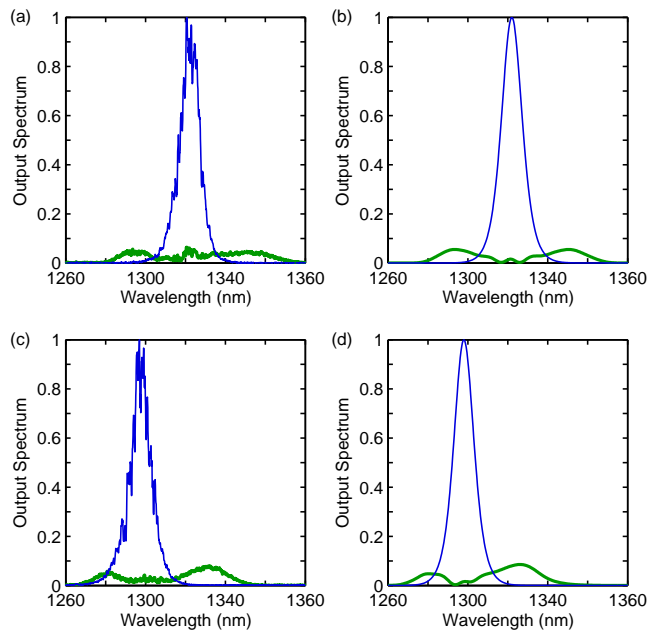}
\caption{Output signal spectrum without (thin blue) and with (thick green) the pump pulse. Experiments at the GVM wavelength of 1322\,nm for the signal (a), and at 1298\,nm (c); (b) and (d), corresponding simulations by means of the CGNLSE. The spectra are normalized to the peak spectral density without the pump. The simulation parameters are: pump peak power $P_p = 3$\,W and delay = 0\,fs (b); and $P_p=$2.15\,W and delay = -100\,fs (d). The collected output average power is measured to -31\,dBm (a) and -33\,dBm (c).}
\label{fig:SpecOut}
\end{figure}

The recorded output spectra with and without the pump are shown in Fig.~\ref{fig:SpecOut}\,(a) and (c). We can readily notice that the interaction with the pump pulse dramatically decreases the output signal power, showing that the cross-TPA is quite large. Moreover, the signal spectral shape is distorted because of cross-phase modulation with the pump. For a signal at the group velocity match (GVM) wavelength of 1322\,nm [see in Fig.~\ref{fig:SpecOut}\,(a)] and a zero delay between the two input pulses, the shape of the spectrum stays symmetric around the signal carrier frequency. The spectrum does not remain symmetric and a large part of the spectrum is located above the GVM wavelength for a delayed input signal at 1298\,nm [see in Fig.~\ref{fig:SpecOut}(c)]. This frequency shift is in overall agreement with Fig.~\ref{fig:Dispersion_curve} that predicts a shifted wavelength of 1350\,nm. The discrepancy can likely be explained by the pump power decrease due to TPA and the relatively long pulses. Note that the initial delay between the pump and the signal has been experimentally set to maximize this frequency shift.  

The efficiency of the cross-absorption was then studied as a function of the initial delay between the pump and the signal pulses. The electronic ND-TPA is an extremely fast nonlinear effect, contrary to the signal absorption mediated by photo-induced free carriers generated by self- or cross-TPA. It is thus expected the signal transmission to be highly sensitive to the pump-signal delay. Fig.~\ref{fig:trans}(a) and (b) show the transmitted signal energy, normalized with respect to the maximum transmission, as a function of this delay for both 1322\,nm and 1298\,nm signal pulse. For the group-velocity matched case, two different pump powers were considered. These results clearly confirm the electronic origin of the nonlinear absorption, with a fast response time $\ll 100$\,fs. Moreover, as there is no significant transmission difference between large positive and negative delays (or equivalently $\approx 12$\,ns delay from the preceeding pump pulse), the free carriers do not play a role in the experiment, contrary to\,\cite{zhang_non-degenerate_2015, suda_femtosecond_2009}. Note that the recorded curves cannot be considered as cross correlation traces between the pump and the signal because both pulses are distorted when propagating in the dispersive nonlinear waveguide.   

\section{\label{sec:level3} Modeling}

In order to compare our experimental results with theory, we first consider the coupled equation model for describing the interaction between the two copropagating pulses~\cite{lin_nonlinear_2007}. These equations have been derived assuming that the cross-TPA coefficient $\beta_{TPA}(\omega_1, \omega_2)$ and the cross-phase modulation coefficient $n_{2}(\omega_1, \omega_2)$ could be replaced by their degenerate value at the mean frequency. Previous works have demonstrated that the Raman effect is negligible when dealing with short pulses in silicon\,\cite{yin_soliton_2007,leo_dispersive_2014}. This effect has not been taken into account in the equation model which thus reads:
\begin{eqnarray}\label{Eq_CGNLSE-1}
\frac{\partial E_{p}(z,t)}{\partial z} - \mathcal{F}^{-1}\left[iD_p(\omega)\tilde{E}_{p}(z,\omega)\right] +\frac{1}{2}\left(\alpha_{p} + \sigma N_c\right)E_{p} \nonumber\\  
+ ik_{0p}k_cN_cE_{p}- i\left(\gamma_{pp}|E_{p}|^2+2\gamma_{ps}|E_{s}|^2\right)E_{p}  = 0,\nonumber\\
\\
\label{Eq_CGNLSE-2}
\frac{\partial E_{s}(z,t)}{\partial z} - \mathcal{F}^{-1}\left[iD_s(\omega)\tilde{E}_{s}(z,\omega)\right] +\frac{1}{2}\left(\alpha_{s} + \sigma N_c\right)E_{s} \nonumber\\  
+ ik_{0s}k_cN_cE_{s}- i\left(\gamma_{ss}|E_{s}|^2+2\gamma_{sp}|E_{p}|^2\right)E_{s}  = 0,\nonumber\\
\end{eqnarray}
where $E_{p,s}(z,t)$ describe the slowly varying envelope of the pump and signal fields as a function of the propagation distance $z$ and time $t$, and
$\tilde{E}(z,\omega)=\mathcal{F}\left[E(z,t)\right]$ is the Fourier transform of the field.  

The dispersion operator at the pump frequency is defined as
\begin{equation}\label{Eq_D}
D_{p}(\omega) = \beta(\omega)-\beta (\omega_{p})-\left.[\partial\beta/\partial\omega\right]|_{\omega_{p}}(\omega-\omega_p),
\end{equation}
where $\beta(\omega)$ is the frequency dependent wavenumber and $\omega_{p}$ denotes the pump frequency. At the signal frequency $\omega_s$, the dispersion operator is $D_{s}(\omega) = D_{p}(\omega-(\omega_s-\omega_p))$. The wavenumber    
$\beta(\omega)$ was numerically computed by use of a mode solver (Lumerical). $\alpha_{p,s}$ characterize the linear loss experienced by the pump and the signal. These two coefficients where experimentally evaluated at 2~dB/cm. 

\begin{figure}[t]
\centering
\includegraphics[width=\linewidth]{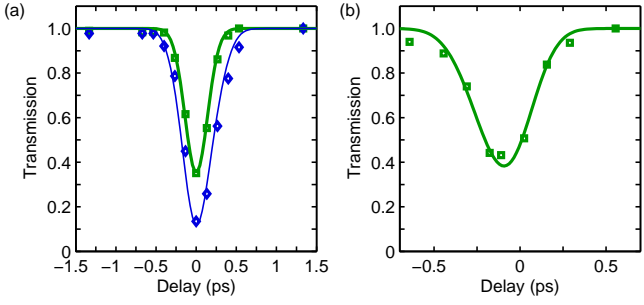}
\caption{Normalized transmission of the signal as a function of the the input signal-pump delay, for a pump wavelength of 1848\,nm. (a) At GVM wavelength (1322\,nm). Experiments at average collected power of -28\,dBm (square) and -35\,dBm (diamond). Simulations for  $P_p = 1.3$\,W (thick green) and 5.2\,W (thin blue). (b) At 1298\,nm signal wavelength. Experiments at an average collected power of -25\,dBm (square) and simulation of the CGNLSE for $P_p = 0.86$\,W.}
\label{fig:trans}
\end{figure}

For the sake of completeness, the effect of photo-generated free carriers of density $N_c$ were taken into account.  
$\sigma N_c$ corresponds to the free carrier induced loss, and the carrier density $N_c$ is calculated by solving the rate equation:
\begin{eqnarray}\label{Eq_carriers}
  \frac{\partial N_\mathrm{c}(z,t)}{\partial t} =
    \frac{Im\left[\gamma_{pp}\right]}{\hbar\omega_p A_{pp}}|E_p(z,t)|^4 + \frac{Im\left[\gamma_{ss}\right]}{\hbar\omega_s A_{pp}}|E_s(z,t)|^4 + \nonumber\\
    4\frac{Im\left[\gamma_{ps}\right]}{\hbar\omega_p A_{ps}}|E_p(z,t)|^2|E_s(z,t)|^2 - \frac{N_\mathrm{c}(z,t)}{\tau_\mathrm{c}}\,.\nonumber\\
\end{eqnarray}

The complex nonlinear parameters $\gamma_{\mu\nu}$ account for the self- and cross- amplitude and phase modulations. They are given by
\begin{eqnarray}\label{def_gamma}
\gamma_{\mu \nu} = \frac{\omega_\mu\eta_{\mu \nu}\overline{n}^2n_2(\overline{\omega})}{c\sqrt{A_{\mu \mu}A_{\nu\nu}}n_\mu n_\nu}
+i\frac{\omega_\mu\eta_{\mu \nu}\overline{n}^2\beta_\mathrm{TPA}(\overline{\omega})}{2\overline{\omega}\sqrt{A_{\mu\mu}A_{\nu\nu}}n_\mu n_\nu},
\end{eqnarray}
where $A_{\nu \nu}$, $n_\nu$ are the effective area related to a third order nonlinear process, and the mode index, at the frequency $\omega_\nu$, calculated with the mode solver. $\overline{n}$ is the mode index at the average frequency $\overline{\omega} = \left(\omega_\mu+\omega_\nu\right)/2$. $n_2$ and $\beta_{TPA}$ are the Kerr nonlinear refractive index and the two-photon absorption coefficients. Finally, $\eta_{\mu\nu}$ is the mode-overlap factor as defined in [\citenum{lin_nonlinear_2007}]. The computed effective area are $A_{pp}=0.241$\,$\mu$m$^2$, $A_{ss}=0.147$\,$\mu$m$^2$, and $A_{ps}=0.176$\,$\mu$m$^2$ and the complex nonlinear coefficients are, based on the data from\,\cite{bristow_two-photon_2007}:\\ 
\\
$\gamma_{pp} = (164+i*14)$\,W$^{-1}$m$^{-1}$, \\
$\gamma_{ss} = (154+i*59)$\,W$^{-1}$m$^{-1}$,\\
$\gamma_{ps} = (97+i*31)$\,W$^{-1}$m$^{-1}$,\\
$\gamma_{sp} = (138+i*45)$\,W$^{-1}$m$^{-1}$.\\

\begin{figure}[tbp]
\centering
\includegraphics[width=\linewidth]{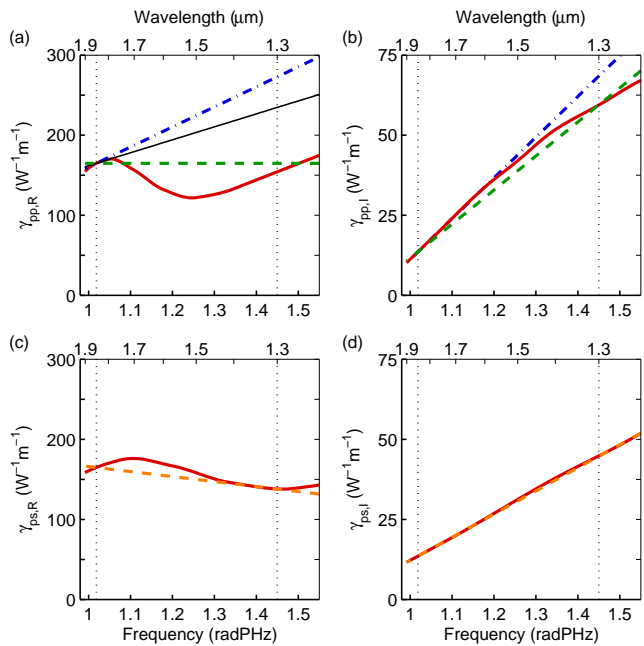}
\caption{Dispersion of the degenerate real (a) and imaginary (b) parts of the nonlinear coefficient $\gamma$ (thick red curve). The vertical dotted lines correspond to the angular frequency of the pump (1.02\,radPHz) and the signal (1.45\,radPHz). The thin black line in the left panel shows the dispersion corresponding to $s_R=1/\omega_p$. The blue (dash-dotted) lines show the first order linear approximation, and the green (dashed) lines correspond to $s_R=0$\,fs and $s_I=7.5$\,fs. Dispersion of the real (c) and imaginary (d) parts of the cross nonlinear coefficient $\gamma_{ps}$ (red curve). The orange (dashed) lines correspond to  $s_R=-0.39$\,fs and $s_I=5.15$\,fs. See also in Fig.~\ref{fig:comp_model}. }
\label{fig:disp_gamma}
\end{figure}

While the real part of $\gamma$ is almost identical for the pump and the signal, the self-TPA coefficient varies by more than a factor of four (see also in Fig. \,\ref{fig:disp_gamma}). As for the cross-TPA coefficient for the signal, it is three times larger than the self-TPA coefficient at the pump wavelength.   

The coupled generalized nonlinear Schr\"{o}dinger equations (CGNLSE) model Eq.~(\ref{Eq_CGNLSE-1})-(\ref{Eq_CGNLSE-2}) is appropriate for describing the two waves interaction since their spectra do not overlap. These set of equations were numerically integrated by a split-step Fourier method, with the input pump power as the only free parameter (all the other parameters have either been computed or accurately been measured). Remarkably, all the experimental results are very well reproduced by numerical simulations. The amplitude and the shape of the transmitted signal agree, both when the signal and the pump propagate with the same group velocity [Fig. ~\ref{fig:SpecOut}(a)-(b)] and when the signal is slightly slower than the pump pulse [Fig.~\ref{fig:SpecOut}(c)-(d)]. There is also a very good agreement for the  transmission of the signal energy as a function of the relative delay between the signal and the pump (see in Fig.~\ref{fig:trans}). These results confirm that a 10 dB ultrafast attenuation of the signal can be obtained with a pump pulse energy as low as 1.2\,pJ. This energy is lower than the ones reported at telecom wavelengths for quasi degenerate TPA in silicon \,\cite{suda_femtosecond_2009} and in amorphous silicon\,\cite{shoji_ultrafast_2010}, thanks to a large cross absorption of the signal but a low TPA at the pump wavelength. This shows the potential of silicon for ultrafast all-optical signal processing with spectrally distant waves. These results also demonstrate that owing to the short pulse duration, the density of photo-generated carriers is too low to significantly impact the cross absorption dynamics, even if the pump photons energy is still above the half-bandgap. The simulations for large delays between the pump and the signal reveal that the free-carrier absorption is only responsible for 2\,\% of the signal attenuation. Finally, the excellent agreement between the theory and the experiment confirms that the simplification of reducing the non-degenerate nonlinear coefficients $n_2(\omega_p,\omega_s)$ and $\beta_{TPA}(\omega_p, \omega_s)$ to their value at the average frequency $\overline{\omega}$ is valid for indirect-gap semiconductors, at least for photon energy in the range $[0.6-0.85]\times E_g$.

Nonlinear propagation of broadband waves in integrated nanophotonic wire waveguides has recently attracted a lot of attention for its potential for efficient generation of  supercontinuum or frequency comb at low power, or for on-chip generation of ultra-short pulses through temporal pulse compression of high-order soliton. To describe the propagation of these waves in nanophotonic waveguides, the generalized nonlinear Schr\"{o}dinger equation (GNLSE) is widely used. It has historically been successfully applied for describing the nonlinear pulse propagation in optical fiber. In these media, for moderate pulse peak power, the nonlinear absorption is very low and thus neglected. The effect of the dispersion of the nonlinear parameters was first discussed in the context of optical fibers and applied to SCG\,\cite{blow_theoretical_1989,karasawa_comparison_2001,dudley_supercontinuum_2006} where a linear dispersion with the frequency, known as the self-steepening term, is considered in the GNLSE. It was later included in models for the simulation of SCG in integrated waveguides made up of various nonlinear media such as chalcogenide\,\cite{lamont_supercontinuum_2008}, silicon\,\cite{lin_nonlinear_2007,leo_dispersive_2014,singh_midinfrared_2015,ishizawa_octave-spanning_2017}, amorphous silicon\,\cite{leo_generation_2014}, silicon nitride\,\cite{johnson_octave-spanning_2015} or InGaP\,\cite{dave_dispersive-wave-based_2015}, where qualitatively good agreement between the simulations and the experiments were reported. Higher order dispersion terms where considered in  \,\cite{zhang_nonlinear_2014}. Apart from this later work, the dispersion of the nonlinear absorption was not specifically taken into account. However, the dispersion of the TPA is non-negligible and has potentially a larger impact on the nonlinear interaction than the dispersion of the Kerr effect.            

%
The dispersion of the real and the imaginary part of the nonlinear coefficient $\gamma$ for self-induced nonlinear refraction and absorption are plotted in Fig.~\ref{fig:disp_gamma} (see red curves).  For the silicon waveguide under consideration, the dispersion of the TPA coefficient is almost linear. Its relative variation is actually larger than for the nonlinear refraction coefficient which tends to vary around a constant value of about 150\,W$^{-1}$m$^{-1}$ in the wavelength range of interest (1.2--1.9\,$\mu$m). Also plotted in Fig.~\ref{fig:disp_gamma} (a)-(b) are the first order approximations of the dispersion of the nonlinear parameters (blue dash-dotted lines). For the real part of $\gamma$, we thus have in the spectral domain $\gamma_R(\omega) = \gamma_R(\omega_p)\times\left[1+s_R(\omega-\omega_p)\right]$, with $s_R=\gamma_R^{-1}\left[\partial\gamma_R/\partial \omega\right]_{\omega_p}$. Sometimes, the dispersion of $n_2$, $n_{\mu\mu}$  and of the effective area $A_{\mu\mu}$ is neglected\,\citep{leo_generation_2014}, leading to a value of $s_R=1/\omega_p$ (thin black line). We can see that these approximations actually strongly deviate from $\gamma_R(\omega)$ and clearly overestimate the self and the cross-phase modulation at frequencies larger than the pump. A constant $\gamma_R$ approximation would actually be more accurate (dashed green line). On the other hand, not taking the dispersion of the TPA into account leads to an underestimation of the cross-TPA (Note that the linear approximation should break for photon frequencies lower than 0.88\,radPHz (2126\,nm) as the TPA term becomes negative, meanings that signals can be amplified. This is not an issue in this work since the pump does not extend so far.)     

Given the dispersion of $\gamma$; we separately consider the dispersion of the TPA and of the Kerr effect. The GNLSE thus reads:
 %
\begin{eqnarray}\label{GNLS_shock}
\frac{\partial E(z,t)}{\partial z} - \mathcal{F}^{-1}\left[iD(\omega)\tilde{E}(z,\omega)\right] +\frac{\alpha}{2}E \nonumber \\
- i\left[\left(\gamma_{R}(1+is_R\frac{\partial}{\partial t})\right)+ i\left(\gamma_{I}(1+is_I\frac{\partial}{\partial t})\right)\right]|E|^2E=0,\nonumber\\
\end{eqnarray}
where the free carrier effects have been neglected and where $s_{R,I}$ are the characteristic times associated with the linear dispersion of the nonlinear coefficients $\gamma_{R,I}$. 

Eq.~(\ref{GNLS_shock}) is not strictly equivalent to the coupled equations Eq.~(\ref{Eq_CGNLSE-1})-(\ref{Eq_CGNLSE-2}) (See Supplemental Material at [URL will be inserted by publisher] for the relation between the two models). The GNLSE can indeed not properly account for both the self- and the cross- effects since it implies that $\gamma_{ps} = \gamma_{pp}$ and $\gamma_{sp} = \gamma_{ss}$, which is not verified for spectrally distant waves (see for instance in Fig.~\ref{fig:disp_gamma}).   
     
\begin{figure}[t]
\centering
\includegraphics[width=\linewidth]{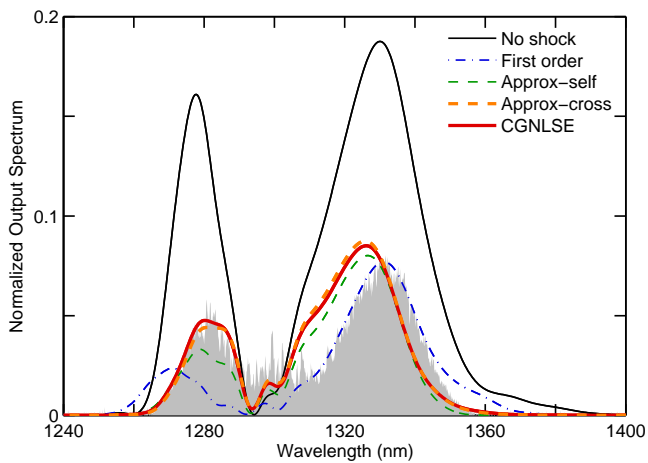}
\caption{Comparison between the different models. No shock: generalized NLS equation (GNLSE) without dispersion of the nonlinearity ($s_R=0$ and $s_I=0$); First order: GNLSE with characteristic times corresponding to the first order approximation of the nonlinear dispersion ($s_R=1.5$\,fs and $s_I=9$\,fs, see also blue lines in Fig.~\ref{fig:disp_gamma}); Approx-self: GNLSE with $s_R=0$\,fs and $s_I=7.5$\,fs, corresponding to the dashed green lines in Fig.~\ref{fig:disp_gamma};  Approx-cross: GNLSE so as the cross-nonlinear coefficents at the signal agree with Eq.~(\ref{Eq_CGNLSE-1})-(\ref{Eq_CGNLSE-2}) ($s_R=-0.39$\,fs and $s_I=5.15$\,fs); CGNLS: simulation with the coupled generalized NLS equations (\ref{Eq_CGNLSE-1})-(\ref{Eq_CGNLSE-2}). The pump is at 1848\,nm and $P_p=2.5$\,W, the input signal wavelength is $\lambda_s=1298$nm. The normalization is the same as in Fig.~\ref{fig:SpecOut}. The shaded gray area show the signal output experimental spectrum. }
\label{fig:comp_model}
\end{figure}

Fig.~\ref{fig:comp_model} shows the simulations of the interaction between the signal and the pump with different characteristic times  $s_{R,I}$ in Eq.~(\ref{GNLS_shock}). These results are compared with the numerical  integration of Eq.~(\ref{Eq_CGNLSE-1})-(\ref{Eq_CGNLSE-2}) considered as a benchmark model for modeling interactions between the two waves. Firstly, we consider the GNLSE without taking the dispersion of the TPA and of the Kerr coefficients into account [$\gamma_R=\Re(\gamma_{pp})$ and $\gamma_I=\Im(\gamma_{pp})$].  The cross-TPA experienced by the signal pulse is lower than its actual value. The output spectral density is clearly overestimated compared with the experiment and the CGNLSE.  A first order approximation of the dispersion of the real and the imaginary parts of $\gamma$ already improves the result. However, the cross-TPA and the cross-phase modulation are now overestimated as seen in Fig.~\ref{fig:disp_gamma}. A better agreement with the CGNLSE is obtained when the dispersion of $\gamma_R$ is neglected ($s_R=0$) while $s_I$ is computed so that $\gamma_I$ takes its correct value  both at the pump and the signal wavelengths, i.e. $s_I=1/\gamma_I(\omega_p)\times\left[\gamma_I(\omega_p)-\gamma_I(\omega_s)\right]/\left[\omega_p-\omega_s\right] = 7.5$\,fs, (see also green curves in Fig.~\ref{fig:disp_gamma}). Nevertheless, there is still a discrepancy between the CGNLSE and the GNLSE because the considered cross absorption coefficient is 59\,W$^{-1}$m$^{-1}$ instead of 45\,W$^{-1}$m$^{-1}$. Actually, as long as the cross-interactions with the pump dominate the nonlinear effects at the signal wavelength, the dispersion of the cross coefficient $\gamma_{sp}$ should be considered. The coefficients $s_R$ and $s_I$ can thus be set such that the cross-phase modulation and the cross-absorption at the signal wavelength take the same values as in Eq.~(\ref{Eq_CGNLSE-2}). We can see in Fig.~\ref{fig:disp_gamma}(c) and (d) that these linear approximations are actually very close to the actual curves on the whole frequency range between the pump and the signal. The two models now give similar results. The small deviations between the two models mainly come from the free carriers effects that are not taken into account in Eq.~(\ref{GNLS_shock}). For historical reasons, only the dispersion of the waveguide Kerr nonlinearity coefficient is usually considered for modeling the nonlinear propagation in integrated seminconductor waveguides. However, as far as silicon is concerned, it is now evident that the dispersion of the TPA coefficient should not be neglected while a constant Kerr coefficient might give better results than a linear approximation around the pump frequency.              

\section{\label{sec:level4} Conclusion}

In this work we have experimentally investigated the nonlinear interaction between two sub-picosecond pulses under cross-TPA and cross-phase modulation in silicon wire waveguides. We have considered an experiment in which the pump is propagating in anomalous dispersion at a wavelength close to the semiconductor half-bandgap (1850\,nm) and a blue-shifted signal at about 1330\,nm. The waveguide dispersion was engineered for the pump and the signal to propagate at almost the same group velocity. The output spectrum was recorded for different input signal wavelengths and different input pump delays. The experiment shows a strong absorption of the signal due to large cross-TPA with the pump. For sub-picosecond pulses, the dynamics of the cross-nonlinear absorption is dominated by the electronic ultrafast two-photon absorption. The free-carrier absorption is negligible as the TPA of the pump is low for wavelengths close or beyond the half-bandgap. A 10~dB amplitude modulation of the transmitted signal has been measured for a pump pulse energy of 1.2\,pJ. These results are better than the one reported in \;\cite{suda_femtosecond_2009} with both the signal and the pump in the telecom wavelength range, because in this latter work both waves experience a large TPA. The experimental results were compared with the numerical integration of a set of two coupled generalized nonlinear Shr\"{o}dinger (NLS) equations. A very good agreement has been obtained for both the transmission curves and the output spectra. From these results, we can conclude that in indirect-gap semiconductors the simplification of computing the cross-nonlinear coefficients from the degenerate $n_2$ and $\beta_2$ values at the average frequency shows excellent results. This could be helpful for optimizing optical functions resorting to cross-interactions between signals at very different wavelengths. The commonly used single generalized NLS equation was also considered for simulating the nonlinear interaction in the waveguide. The dispersion of the Kerr coefficient and of the nonlinear absorption was considered. It appears that the dispersion of the TPA cannot be neglected. Moreover because of the dispersion of the effective area of the mode and of $n_2$, the standard first order approximation of the nonlinear effective Kerr coefficient largely overestimate the cross-phase modulation, while better results are obtained with a constant real part of $\gamma$. The best results are obtained when the coefficients $s_{R,I}$ are such that they properly account for the cross-nonlinear effects. This observation might impact the simulation and the analytical study of supercontinuum and of frequency comb generation in integrated semiconductor structures based on the GNLSE model. A large variation over the spectrum of the nonlinear refraction and of the two and/or three photon absorption, as well as of the effective area, is indeed encountered in these structures.  

\section*{\label{sec:level5} Funding information}

This work is supported by the Belgian Science Policy Office (BELSPO) Interuniversity Attraction Pole (IAP) project Photonics@be and by the Fonds de la Recherche Fondamentale Collective, Grant No. PDR.T.1084.15.

\bigskip

\bibliography{XTPA-paper.bib}

\end{document}